# NetLogo Implementation of an Evacuation Scenario

João Emílio Almeida, Zafeiris Kokkinogenis
Artificial Intelligence and Computer Science
Laboratory (LIACC)
Faculty of Engineering, University of Porto (FEUP)
Porto, Portugal
{ joao.emilio.almeida, pro08017}@fe.up.pt

Rosaldo J. F. Rossetti
Artificial Intelligence and Computer Science
Laboratory (LIACC)
Faculty of Engineering, University of Porto (FEUP)
Porto, Portugal
rossetti@fe.up.pt

*Abstract* - **The problem of evacuating crowded closed spaces, such as discotheques, public exhibition pavilions or concert houses, has become increasingly important and gained attention both from practitioners and from public authorities. A simulation implementation using NetLogo, an agent-based simulation framework that permits the quickly creation of prototypes, is presented. Our aim is to prove that this model developed using NetLogo, albeit simple can be expanded and adapted for fire safety experts test various scenarios and validate the outcome of their design. Some preliminary experiments are carried out, whose results are presented, validated and discussed so as to illustrate their efficiency. Finally, we draw some conclusions and point out ways in which this work can be further extended.**

## I. Introduction

Evacuating large crowds is a challenge under any circumstances. The evacuation from large facilities during an emergency or disaster is a much more complex task because chaos and panic add to highly density populations even more uncertainty and stress. Studying crowd behaviour in emergency situations is difficult since it often requires exposing real people to the actual, possibly dangerous, environment. Fire drills (figure 1) are possible approaches to study this phenomenon. However, as they hardly recreate the truly panic conditions, people tend not to take it seriously. A simulation tool could serve as a viable alternative.

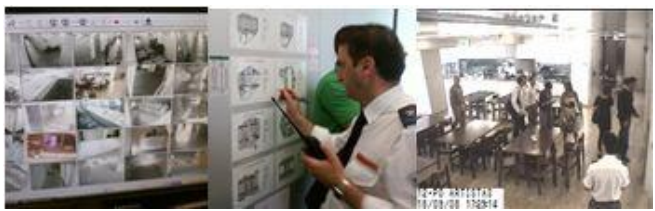

Figure 1. Fire drills

Crowds and pedestrians have been empirically studied for the past decades [1, 2]. The evaluation methods applied were based on direct observation, photographs, and time-lapse films. Apart from behavioural investigations, the main goal of these studies was to develop computer animated realistic applications, for the game industry, design elements of pedestrian facilities, or planning guidelines for architectural building and urban design.

In their common environment pedestrians tend to show some basic attributes. For example people always try to find the shortest and easiest way to reach their destination. If possible they avoid detours, even if the shortest way is crowded. The basic principle is the "least effort principle", which means everyone tries to reach their goal as fast as possible spending the least amount of energy and time. Observations made for crowds in emergency situations feature typically the same patterns. As people try to leave the building as fast as possible, the desired velocity increases which leads to some characteristic formations. As nervousness increases there is less concern about comfort zone and finding the most convenient and shortest way. It is observable, for example, that if people have to leave a building in an emergency situation and they don't know the structure of the building well enough, they would run for the exit they used as an entrance, even if other exits might be easier to reach or even safer [1].

In this paper we focus our attention on the evaluation of to the architectural layout of a closed public space as it can be a museum, cinema, discotheque or concert hall. Thus are not taken into consideration typical behavioral theories we find in crowd simulations. For this work we used the NetLogo [3] agent-based modelling and simulation platform to rapidly prototype simple, yet realistic, "what-if" evacuation scenarios.

The rest of the paper is organized as it follows: in the section two some crowd and evacuation simulation related works are presented. In the section 3 we present the model implementation, in section 4 and 5 the model validation and the experimental result discussion are shown respectively. Finally in section 6 we give space to some final remarks and future works.

## II. Related Works

There are three main reasons for developing computer simulations for crowd behaviours: firstly, to test scientific theories and hypotheses; secondly, to test design strategies;

and finally, to recreate the phenomena about which to theorize [4]. Computer models for emergency and evacuation situations have been developed and much research on panics has been of empirical nature and carried out by researchers from the social sciences [1, 2, 5, 6, and 7].

Santos et al. [8] offer a critical review of selected simulation models of evacuation. Also, authors have identified social sciences approaches that could improve contemporary simulation models. They argue that social sciences could provide important new directions to simulation models of emergency evacuations. Helbing et al. [1, 2] developed a continuous pedestrian model based on plausible interactions, and pointed out that pedestrian dynamics shows various collective phenomena, such as arching, clogging and herding [5]. According to their findings, every simulation model should reproduce such behaviours in order to be rather realistic. They believe that the above models can serve as an example linking collective behaviour as a phenomenon of mass psychology (from the socio-psychological perspective) to the view of an emergent collective pattern of motion (from the perspective of physics). Their simulations suggest that the optimal behaviour in escape situations is a suitable mixture of individualistic and herding behaviour. Kaup et al. [9] present a simulation model for emergency planning and crowd management purposes. Other simulation studies, combined with optimization algorithms, aim at improving the evacuation efficiency. This can be achieved in terms of evacuation times, assessment and analysis of evacuation plans, as well as routes/path optimizations. Coelho [10] has deeply studied the mathematical and analytical models, comparing and analyzing them on a systematic basis. In Klüpfel et al. [11], the analysis of evacuation processes on-board is taken into consideration. Here authors apply cellular automata to reproduce crowd motion for the detection of possible bottleneck situations during the evacuation process. Filippoupolitis et al. [12] address the building evacuation optimization problem. Su et al. [13] developed a discrete-event computer simulation model for assessing evacuation programs and provide a comprehensive idea of evacuation plans for hospital buildings in the event of a possible bomb threat.

### III. MODEL IMPLEMENTATION

In this paper we describe the use of NetLogo as a rapid prototyping tool for an agent-based evacuation simulation study of a closed space with the purpose to assess the building's architectural layout in terms of evacuation times and occupants exit rate, only. In this section, we present how our problem was modelled and which abstractions were used to achieve it. Furthermore, complexities and constraints inherent to this problem were identified. From that, a simplified model of an abstraction of the application domain was created without losing key aspects, such as certain degrees of realism. Our purpose is to simulate the evacuation of a closed space evaluating the layout of its egresses considering a simple behaviour of the occupants only. The goal is to evaluate and compare scenarios' performance, changing the location and width of the exits using as metric the total evacuation time.

As agents, we defined the occupants of a closed environment randomly distributed. All agents have the same characteristics of adult population. Each agent will head always toward the nearest exit. If other agents occupy the space around him, he will wait for his turn to move on. No overlapping is allowed. This scenario is quite simple but we can still observe some of the typical arching and clogging phenomena. The devised scenario consists on a square configuration of 55 x 55 patches. Considering that each patch has 1 m width, they will have an area of 1 m2. Therefore, the space has 55 m x 55 m = 3025 m2. Occupants are represented as a circle, and only one person per patch is allowed. Thus, the maximum occupancy density is 1 person per meter-square. An instantiation of the model can be seen in figure 2.

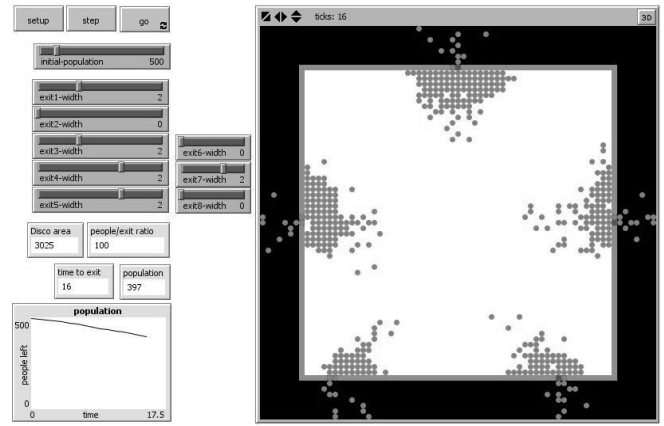

Figure 2. NetLogo model implementation

Patches are in one of three states: inside, outside or wall. Outside patches are black, inside white and the walls are grey. The space limits, or walls, are special patches that cannot be used by occupants. Exits are controlled using "sliders", and their width might vary from 0 to 1, 2 or 3, depending on the situations. To make this space more realistic, the exits were located following the Portuguese law (detailed in the next section), requiring a certain number and distance between them. Some exits were placed having their width controlled using the correspondent slider. Each position corresponds to a meter. So, an exit with width of 2 units means that the door has 2 meter width. If zero means that the exit is closed and nobody can go outside through it. The number of occupants is also variable. For the sake of simplification, the closed space is represented with no obstacles inside. This happens sometimes in the real life, when rock concerts happen in gymnasiums.

When the simulation setup starts, occupants are distributed inside the area of interest. After pressing the "go" bottom, meaning that a complete evacuation order was given, for instance due to a fire or bomb threat, all occupants start moving towards the nearest exit. The simulation ends when all occupants exit the space. The movement of each occupant is considered to be 1 m/s. If each iteration (or tick) lasts a second, then the occupants shall move from one patch to another in each tick.

## IV. VALIDATION METHODOLOGY

To the validation of the model, mathematical and empirical knowledge were used. A comparison using common analytical models (such as Van Bogaert method or the Predetchenky & Milinskii formulae) was carried out to validate the data obtained through NetLogo, as explained below.

### A. The Van Bogaert analytical method

The Van Bogaert [10] approach assumes that horizontal movement speed is 1.5 m/s. Total evacuation time depends on two main factors: total number of occupants and the flow of the exits. The maximum time to evacuate the space is given by:

$$T_{max} = S \times I \times F_d \times H \times R \times 300 \text{ [s]} \quad (1)$$

The components of (1) are coefficients found empirically, representing the various factors that influence the final result. $S$ depends on the area (in this case S=3), I reflects the building compartment (in this case is none, thus I=0.75), $F_d$ is the density factor ($F_d = 0.36$ for maximum density), H is related with the building height ($H$=1 since it is a one storey building), and $R$ is risk type ($R$=1 for ordinary risks). Replacing the coefficients in (1), we have the following expression:

$$T_{max} = 3 \times 0.75 \times 0.36 \times 1 \times 1 \times 300 = 243 \text{ [s]}$$

So, according this method, the total evacuation time should be 4 minutes (or to be more precise, 243 seconds).

### B. The Predetchenky & Milinskii method

Predetchenky & Milinskii [10, 14] developed analytical methods to calculate the evacuation speed of crowd. They were the ground basis to yield the following formulae:

$$V = (112 D_a^4 - 380 D_a^3 + 434 D_a^2 - 217 D_a + 5\,7)/60 \quad (2)$$

$$V_{VE} = V_V \times (1,49 - 0,36 D_a) \quad (3)$$

Where V (2) is the walking velocity [m/s] in normal situation and $V_{VE}$ (2) is the horizontal velocity [m/s] in emergency situation. $D_a$ corresponds to the dimensionless density given by:

$$D_a = (N \times P_h)/A \quad (4)$$

In (4), $N$ is the total number of persons, $P_h$ the area occupied by a person (typically is 0.125 m2) and $A$ is the area of the space in m2 (3025 m2 in the example). Replacing the values in the equations, first in (4) $D_a = 0.041322$, then in (2) $V = 0.81246$ m/s and $V_{VE} = 1.19848$ m/s.

### C. The Portuguese fire code

The Portuguese Fire Code considers, regarding the exits: a) the maximum distance from any given point to the nearest exit; b) the number of exits; c) the sum of the exits width. For this model we assume that the minimum distance requirement to the exits was valid. The number of exits and their respective width will depend on the expected number of occupants. Passage Unit (PU) concept is basically defined through empirical studies and determines the minimum width through which a certain number of egresses can leave the space safely. It is considered that for each group of 100 occupants one PU is required. In Portuguese Fire Code, the first PU is 0.9m, second 1.4m and onwards is $n \times 0.6$m where n is the number of PU considered. For the present study, it was assumed that each PU corresponds to 1m width. Another constraint regards the distribution of the exits that must be independent. In other words, exits considered independent if they are located from each other a distance far enough to form an angle of no less than 45º to the user observing both exits.

For the given example, having 1,000 occupants, the Portuguese Fire Code requires a minimum of 5 independent exits, with 2 PU each, resulting in a total of 10 PU (10 x 0.6m = 6m width). In our example this width is 10m.

## V. EXPERIMENTAL RESULTS AND DISCUSSION

After setting up the model, some experiments were carried out. Taking advantage of the random location of occupants, at least three runs were performed for each scenario. The number of occupants was fixed to 1,000. The exits varied from 1 to 8, as well as varied their width PU. The results are presented in Table I.

Table I. Results of experiments using Netlogo

| exits | PU width | total width | runs (seconds to exit) #1 | #2 | #3 | exits parameters 1 | 2 | 3 | 4 | 5 | 6 | 7 | 8 | remarks |
|---|---|---|---|---|---|---|---|---|---|---|---|---|---|---|
| 1 | 1 | 1 | 1261 | 1273 | 1258 | 0 | 1 | 0 | 0 | 0 | 0 | 0 | 0 | only 1 exit with 1PU |
| 2 | 1 | 2 | 670 | 666 | 656 | 0 | 1 | 0 | 0 | 0 | 0 | 1 | 0 | 2 opposit exits 1 PU each |
| 2 | 2 | 4 | 364 | 366 | 368 | 0 | 2 | 0 | 0 | 0 | 0 | 2 | 0 | 2 opposit exits 2 PU each |
| 4 | 1 | 4 | 358 | 340 | 356 | 0 | 1 | 0 | 1 | 1 | 0 | 1 | 0 | 4 opposit exits 1 PU each |
| 4 | 2 | 8 | 202 | 190 | 197 | 0 | 2 | 0 | 2 | 2 | 0 | 2 | 0 | 4 exits with 2UP each |
| 5 | 2 | 10 | 187 | 179 | 183 | 2 | 0 | 2 | 2 | 2 | 0 | 2 | 0 | 5 exits with 2PU each |
| 6 | 2 | 12 | 156 | 170 | 158 | 2 | 0 | 2 | 2 | 2 | 0 | 2 | 2 | 6 exits with 2UP each |
| 8 | 2 | 16 | 136 | 143 | 142 | 2 | 2 | 2 | 2 | 2 | 2 | 2 | 2 | 8 exits with 2UP each |

For each of the eight possible exits the parameter (figure 3) value varies from 0 (meaning the door is closed) to 2, the PU value.

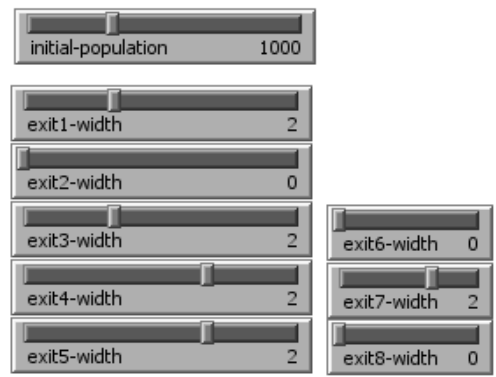

Figure 3. Model parameters

These results show that increasing the number of exits reduces total evacuation time. There is a relation between the exits and width, even when the total width is the same, more exits result in a shorter evacuation time. That is something someone could guess by intuition. Nonetheless, using simulations such as the one presented in this work, such an assumption can be demonstrated and confirmed.

Considering the scenario that better represents the Portuguese Fire Codes, with 5 exits having 2 PU of width each, and running simulations three times, the results obtained were 187, 179, 183, seconds respectively (figure 4). Comparing the results with the validation reference methods, described previously, the simulation speed was 1 m/s, whereas the other methods use 1.5 m/s (i.e. in Van Bogaert) and 1.19848 m/s (i.e. in Predetchenkii & Milinskii). The total evacuation time, according to the Van Bogaert analytical method should be less than 243 seconds. Therefore, considering the results for the selected scenario are always smaller than that reference value, even if occupants speed is slower than the other methods suggest. So the Portuguese Fire Code requirements give better results than the other methods.

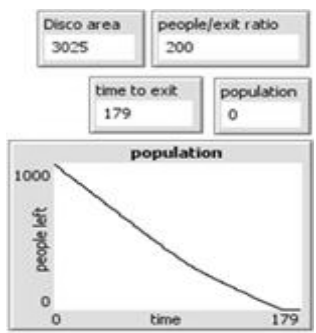

Figure 4. Simulation results

## VI. Conclusions and Future Work

In this paper we presented an evacuation model implemented in the NetLogo platform. Although the model may appear to be simplistic, its conceptualisation encompasses some aspects of the observed system in the real world. As such, we are able to observe typical emerging behaviour patterns during evacuation situations, namely arching, clogging and other herding phenomena. The proposed model presents a satisfactory degree of validity within a plausible range of situations. In order for the model to achieve higher degree of validity, more realistic features should be incorporated, both in the agent behaviour model as well as in the representation of the environment. Regarding the environment, different layouts should be taken into account (including rooms, hallways and obstacles). It will be very desirable the implementation of a realistic fire/smoke diffusion model as well, where heat, visibility and smoke can be accounted for. This ultimately would influence occupant's decisions as their cognitive abilities would be also affected. Another important issue is the development of a social force and psychological model to represent the agent internal state providing a more realistic simulation of the human decision-making mechanisms, as was suggested in [1,2]. These features will give a very reliable and true perspective of an emergency case in a closed space offering to the designers the possibility to better understand the dynamics that take place in such situations. From what we have experienced, our approach is feasible and may support more complex representations of this kind of applications, proving to be an important tool for designing and validation of evacuation plans and strategies.

Acknowledgment

This project has been partially supported by FCT (Fundação para a Ciência e a Tecnologia), the Portuguese Agency for R&D, under grants SFRH/BD/72946/2010, SFRH/BD/67202/2009